\documentclass[fleqn,twoside]{article}
\usepackage{espcrc2}

\usepackage{graphicx}

\def\simge{%  ``less than about'' symbol
    \mathrel{\rlap{\raise 0.511ex
        \hbox{$>$}}{\lower 0.511ex \hbox{$\sim$}}}}

\title{Excited nucleon spectrum from lattice QCD \\
with maximum entropy method}

\author{K.~Sasaki\address[TOKYO]
        {Department of Physics, University of Tokyo, Tokyo 113-0033, Japan}, 
        S.~Sasaki\addressmark, 
        T.~Hatsuda\addressmark and
        M.~Asakawa\address
        {Department of Physics, Kyoto University, Kyoto 606-8502, Japan}}

\date{}

\begin{document}

\begin{abstract} 
We study excited states of the nucleon in quenched lattice QCD
with the spectral analysis using the maximum entropy method.
Our simulations are performed on three lattice sizes
$16^3\times 32$, $24^3\times 32$ and $32^3\times 32$,
at $\beta=6.0$ to address the finite volume issue.
We find  a significant finite volume effect on
the mass of the Roper resonance for light quark masses.
After removing this systematic error, its mass
becomes considerably reduced toward
the direction to solve the level order puzzle 
between  the Roper resonance $N'(1440)$ and 
the negative-parity nucleon $N^*(1535)$. 
\end{abstract}

\maketitle

A particular interest in the excited nucleon spectra
 is the level order of the positive-parity excited nucleon
 $N'(1440)$ (the Roper resonance) and the 
negative-parity nucleon $N^*(1535)$. 
 This pattern of the level order is also seen
universally in the excited states of flavor octet and decuplet baryons.
The quark confining models such as the harmonic oscillator 
quark model or  the MIT bag model 
have difficulties in reproducing the Roper as the first
 excited state of the nucleon  \cite{review}.
 
First systematic lattice QCD calculation  for the nucleon excited states 
in both parity channels  shows that the wrong ordering 
between $N'$ and $N^*$  also takes place at least for  {\it 
heavy-quark} masses~\cite{Sasaki:2001nf}.
After this work, many lattice calculations confirmed 
this puzzle, which are summarized in Ref. \cite{Sasaki:PTP}. 
 
Recently,  it was pointed out that
the  finite lattice volume affects the structure of the  
Roper resonance considerably for light quark 
masses~\cite{Sasaki:PTP,S.Sasaki02}.  
If the  ``wave function'' of a quark  inside baryon  is
squeezed due to the small volume,
the kinetic energy of internal quarks increases and  
the total energy of the bound state should be pushed up.
Such effect is expected to become serious for 
radially excited states as compared to the ground state.
In the present report,  we show that the Roper resonance 
particularly receives large finite volume effect: its mass
after removing the finite volume effect moves toward
the direction to solve the level order puzzle.
 
In the numerical simulations, we generate quenched QCD configurations 
with standard single-plaquette Wilson action mainly at $\beta=6.0$ 
($a^{-1}\sim 2$GeV) for three different lattices,  
$16^3\times 32$ (444 conf.),
$24^3\times 32$ (350 conf.),  and $32^3\times 32$ (200 conf.) 
to  study   finite volume effect. 
The quark propagators are computed using the Wilson fermion action 
at four values of the hopping parameter, which cover the range 
$M_\pi/M_\rho \simeq 0.69-0.92$. 
We adopt the conventional interpolating operator 
$\epsilon_{abc}(u_a^TC\gamma_5d_b)u_c$ to create the positive parity
states of the nucleon.

To extract the mass of the excited nucleon,
we utilize the maximum entropy method (MEM) \cite{Y.Nakahara99}
which can reconstruct  the 
spectral function (SPF) $A(\omega)$
from  given Monte Carlo data of two point hadron correlator 
$G(t)=\int d\omega A(\omega)\exp(-\omega t)$. 
Details of the MEM analysis can be found in~\cite{{S.Sasaki02},{Y.Nakahara99}}.

%
% FIG
%
\begin{figure}[htp]
  \begin{center}
  \includegraphics[height=45mm]{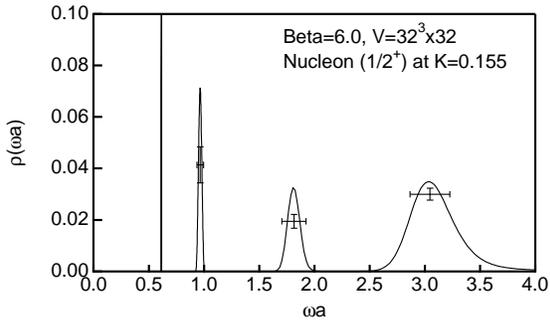}
  \end{center}
  \vspace{-30pt}
  \caption{
 The dimensionless SPF $\rho(\omega)=A(\omega)\omega^{5}$ 
 in the nucleon channel as function of the frequency $\omega$ in lattice units. 
  }
  \vspace{-20pt}
  \label{fig:SPF}
\end{figure}

First we show dimensionless SPF of the nucleon for the largest lattice size
at the lightest quark mass as a typical example in Fig.\ref{fig:SPF}.
There are two sharp peaks and two large bumps. 
The crosses on each peak or bump represent the statistical significance of SPF 
obtained  by the MEM~\cite{Y.Nakahara99}. 
We remark that the first two peaks are physical states but two large bumps 
are unphysical states. This is confirmed by the 
observation that those bumps appear 
at the same dimensionless frequency $\omega a$ in simulations
with different lattice spacings. It indicates that those states become 
infinitely heavy and decouple from physical states in the continuum limit.
The similar observation in the mesonic case has been reported in Ref.~\cite{Yamazaki:2001er}.

\begin{figure}[t]
  \begin{center}
  \includegraphics[height=60mm]{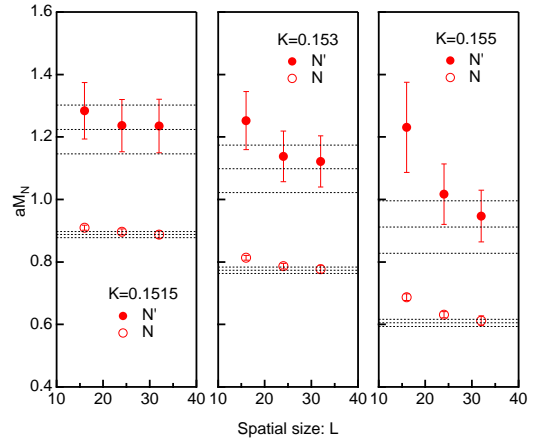}
  \end{center}
  \vspace{-30pt}
  \caption{Finite volume effect on the positive-parity nucleon channel.
  Masses of the ground state (open symbols) 
  and the first excited state  (filled symbols) for three lightest quark masses
  as a function of the spatial lattice size $L$.}
  \vspace{-20pt}
  \label{fig:volume}
\end{figure}

In Fig.~\ref{fig:volume} we plot the masses of the ground state (open circles) 
and the first excited state (filled circles), which correspond to the peak positions 
of first two peaks in SPF, as functions of the spatial lattice size $L$.
The quoted errors are estimated by the jackknife method. 
Observed finite volume effect is clearly sizable
for light quarks (the right panel) and its
effect is larger for the excited state than the ground state.
Dashed horizontal lines in Fig.\ref{fig:volume}
express the masses in the infinite volume limit
and their errors guided by a formula 
$M_{L}=M_{\infty}+c L^{-3}$~\cite{M.Fukugita92}. 
For the lightest quark mass in our calculation, $L$ 
should be as large as 30 to remove the 
finite volume effect. 
This corresponds to $La\approx 3.0 $ fm in the physical unit.

%
% FIG
%
%

%
% FIG
%
\begin{figure}[t]
  \begin{center}
  \includegraphics[height=70mm]{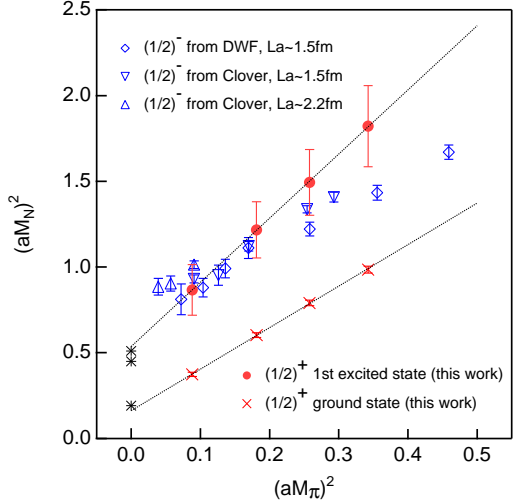}
  \end{center}
  \vspace{-30pt}
  \caption{
    Ground and excited nucleon spectra as a function of the 
    the pion mass squared. The corresponding experimental
    values for $N(940)$, $N'(1440)$ and $N^*(1535)$ are marked with
    lower, middle and upper stars.}
   \vspace{-20pt}
  \label{fig:switching}
\end{figure}

To see whether the level switching 
between $N'$ (the Roper) and $N^*$
(the negative parity nucleon)
may possibly occur, the squared masses of these 
resonances are plotted in Fig.\ref{fig:switching} as 
a function of the pion mass squared.
The filled circles are our results of the Roper in
the infinite volume limit, while the open
squares and triangles are the data at the same lattice spacing
for the negative parity nucleon taken from Refs.~\cite{Sasaki:2001nf,Gockeler:2001db} 
for $La\approx1.5$ fm and $2.2$ fm.
Although the negative parity state is lighter for heavy quark
masses, there is a tendency that the $N'$-$N^*$ level splitting becomes smaller 
or even inverted as the quark mass decreases.
Note here that the finite volume effect is not significant 
for  the negative parity state~\cite{Gockeler:2001db} as can be seen by the 
comparison among the open symbols in Fig.\ref{fig:switching}.
This is in contrast to the significant finite volume effect
on the Roper resonance
shown in Fig.\ref{fig:volume} and in Refs.~\cite{Sasaki:PTP,S.Sasaki02}.

Taking a simple linear extrapolation toward the chiral limit in Fig.\ref{fig:switching},
we find $M_{N'}=0.73(14)$ in lattice units. Our $N'$ mass is evidently smaller than
the previously published results for the $N^*$ at the same lattice spacing;
$M_{N^*}=0.85(5)$~\cite{Sasaki:2001nf} from domain wall fermions and
$M_{N^*}=0.89(2)$~\cite{Gockeler:2001db} from clover fermions.
 
Next we demonstrate the importance of the 
large spatial volume in a different way. We calculated the quark mass
dependence of the mass splitting between the ground state
and radially excited state. From the phenomenological point of view, such a mass splitting 
is almost independent of the quark mass as pointed out in Ref. \cite{Sasaki:PTP}.
%
% FIG
%
\begin{figure}[ht]
  \begin{center}
  \includegraphics[height=40mm]{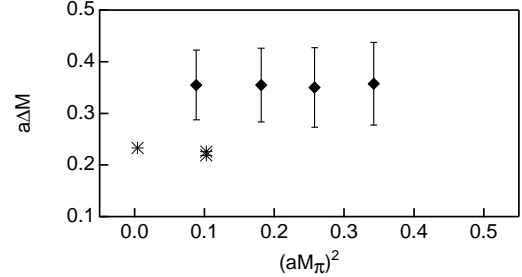}
  \end{center}
    \vspace{-30pt}
  \caption{
  The splitting of 1st and 2nd peaks of the
  spectral function as a function of the pion mass squared.
  The corresponding experimental values for $M_{_{N(1440)}}-M_{_{N(940)}}$ and for
  the strange sector, $M_{_{\Sigma(1660)}}-M_{_{\Sigma(1190)}}$ and 
  $M_{_{\Lambda(1600)}}-M_{_{\Lambda(1115)}}$, are marked as stars.}
  \vspace{-15pt}
  \label{fig:mass_dep}
\end{figure}
Shown in Fig.\ref{fig:mass_dep} is our  
lattice results of the $N$-$N'$ mass splitting  for the largest size lattice $La= 3$ fm. 
As is evident from the figure, the mass splittings 
are independent of the quark mass within
the statistical error bars. 
This result is different from those shown in Ref.~\cite{Dong:2003zf}.

%%%%%%%%%%%%%%%%%%%%%%%%%%%%%%%%%%%%%%%
In summary,
we studied the  excited states of the nucleon with the use of  MEM.
On the basis of a systematic analysis with three different lattice volumes
($La = 1.5, 2.2$ and 3.0 fm),
we confirmed our previous finding that
the Roper resonance receives large finite volume effect
for small quark masses.  We found that, for enough
large spatial size ($\simge3.0 {\rm fm}$), the mass splitting of the 
ground and first excited states of the nucleon
is independent of the quark mass. This is
consistent with the experimental mass splittings
of baryons and mesons for wide range of quark masses \cite{Sasaki:PTP}.
By removing the finite volume effect, the mass of the Roper
resonance becomes considerably reduced and  
the level switching  between the Roper and the 
negative parity resonance may take place for
sufficiently small quark masses.
%~\cite{Dong:2003zf}.  

Needless to say, further study is required to check other systematic 
error stemming from a discretization artifact. Additive simulations at 
different lattice spacings are under way.

We appreciate A. Nakamura and C. Nonaka for helping us 
develop codes for our %lattice QCD 
simulations from their open
source codes (Lattice Tool Kit~\cite{LTK}).
This work is supported by the Supercomputer Project No. 85 (FY2002) and
No. 102 (FY2003) of High Energy Accelerator Research Organization (KEK).
S.S. thanks for the supper by JSPS Grand-in-Aid
for Encouragement of Young Scientists (No.15740137).

%%%%%%%%%%%%%%% References %%%%%%%%%%%%%%%%%%%%


\begin{thebibliography}{9}
\bibitem{review}
For a recent review,
S. Capstick and  W. Roberts, \protect{nucl-th/0008028}. 
%%CITATION = NUCL-TH 0008028;%%
  
\bibitem{Sasaki:2001nf}
S.~Sasaki, T.~Blum and S.~Ohta,  Phys. Rev. {\bf D65} (2002) 074503;
%%CITATION = HEP-LAT 0102010;%%
%\bibitem{Sasaki:1999yh}
S.~Sasaki  
Nucl.\ Phys.\ Proc.\ Suppl.\  \textbf{83} (2000) 206;
\protect{hep-ph/0004252}.
%%CITATION = HEP-LAT 9909093;%%
%%CITATION = HEP-PH 0004252;%%

\bibitem{Sasaki:PTP}
S.~Sasaki,
Prog. Theor. Phys. Suppl. {\bf 151} (2003);  \protect{nucl-th/0305014}.
%%CITATION = NUCL-TH 0305014;%%


\bibitem{S.Sasaki02}
S.~Sasaki, K.~Sasaki, T.~Hatsuda and M.~Asakawa,
\protect{hep-lat/0209059}.
%%CITATION = HEP-LAT 0209059;%%

\bibitem{Y.Nakahara99}
Y.~Nakahara, M.~Asakawa and T.~Hatsuda, 
Phys. Rev. {\bf D60} (1999) 091503;
%%CITATION = HEP-LAT 9905034;%%
M.~Asakawa, T.~Hatsuda and Y.~Nakahara,
Prog. Part. Nucl. Phys. {\bf 46} (2001) 459.
%%CITATION = HEP-LAT 0011040;%%


\bibitem{Yamazaki:2001er}
T. Yamazaki {\it et al.} [CP-PACS Collaboration],
Phys. Rev. {\bf D65} (2002) 014501.
%%CITATION = HEP-LAT 0105030;%%

 
\bibitem{M.Fukugita92}
M.~Fukugita {\it et al.}, Phys. Lett. {\bf B294} (1992) 380. 
%%CITATION = PHLTA,B294,380;%%

\bibitem{Gockeler:2001db}
M.~G{\" o}ckeler  {\it et al.} [QCDSF-UKQCD-LHPC Collaboration],
Phys. Lett. {\bf B532} (2002) 63. 
%%CITATION = HEP-LAT 0106022;%%

\bibitem{Dong:2003zf}
S.~J.~Dong {\it et al.}, \protect{hep-ph/0306199}.
%%CITATION = HEP-PH 0306199;%%

\bibitem{LTK}
S.~Choe {\it et al.}, Nucl.\ Phys.\ Proc.\ Suppl.\  \textbf{106} (2002) 1037. 
%%CITATION = NUPHZ,106,1037;%%
\end{thebibliography}
\end{document}